# Heterodyne dispersive cavity ring-down spectroscopy exploiting eigenmode frequencies for high-fidelity measurements

Agata Cygan, Szymon Wójtewicz, Hubert Jóźwiak, Grzegorz Kowzan, Nikodem Stolarczyk, Katarzyna Bielska, Piotr Wcisło, Roman Ciuryło, Daniel Lisak

*Institute of Physics, Faculty of Physics, Astronomy and Informatics, Nicolaus Copernicus University in Toruń, Grudziadzka 5, 87-100 Torun, Poland*

Measuring low light absorption with combined uncertainty < 1‰ is crucial in a wide range of applications. Popular cavity ring-down spectroscopy can provide ultra-high precision, below 0.01‰, but its accuracy is strongly dependent on the measurement capabilities of the detection system and typically is about 10‰. Here, we exploit the optical frequency information carried by the ring-down cavity electromagnetic field, not explored in conventional CRDS, for high-fidelity spectroscopy. Instead of measuring only the decaying light intensity, we perform heterodyne detection of ring-downs followed by Fourier analysis to provide exact frequencies of optical cavity modes and a dispersive spectrum of a gas sample from them. This approach is insensitive to inaccuracies in light intensity measurements and eliminates the problem of detector band nonlinearity, the main cause of measurement error in traditional CRDS. Using the CO and HD line intensities as examples, we demonstrate the sub-‰ accuracy of our method, confirmed by the best *ab initio* results, and the long-term repeatability of our dispersion measurements at $10^{-4}$ level. Such results have not been achieved in optical spectroscopy before. The high accuracy of the presented method indicates its potential in atmospheric studies, isotope ratio metrology, thermometry, and the establishment of primary gas standards.

The challenge of measuring the shape and intensity of spectral lines with a relative accuracy of $10^{-3}$ and better is highlighted in numerous scientific, industrial, and metrological applications using sensitive optical spectroscopy. Regarding the effect of global warming, changes in the Earth's climate are expected to impact the capacity of natural repositories of anthropogenic greenhouse gases (GHG), which will generate a feedback response to climate change[1]. To predict the evolution of these changes, the location of regional sources and sinks of GHG is essential. Spectroscopic retrieval models must exhibit sensitivity to changes in their concentration at the permille level, necessitating a laboratory spectra accuracy of at least an equivalent magnitude[2]. Any systematic errors are of great concern because they introduce regional bias that imitates sources and sinks of GHG. Particular attention is also required in measurements of the stable isotope ratio, as repeatability is compromised over time due to the aging of reference materials[3]. A promising approach involves the spectroscopic measurement of the absolute isotope ratio from the ratio of the line intensities of these isotopes. This method has recently demonstrated[4] a relative combined measurement uncertainty at the sub-‰ level, showing good agreement with the results obtained from other methods. However, to measure line intensities with such high accuracy, careful calibration of the light intensity detection system is necessary[5]. Accurate measurements of the line intensity ratios are also the basis for the new concept of optical primary thermometry[6]. With the current standard of using first principles to define the units of temperature, pressure, and number density[7], optical methods offer promising prospects for realizing new primary gas standards[8]. As molecules interact with electromagnetic radiation, the accurate measurements of the refractive index enable the determination of gas thermodynamic parameters. Cavity-based nitrogen refractometry with a relative precision of $10^{-6}$ holds the greatest potential for realizing an optical primary pressure standard[9,10] to date. On the other hand, individual spectral lines, shaped by molecular interactions, provide molecular selectivity for optical gas standards. Progress in the mutual development of line-shape theory and spectroscopic methods[11] motivates continuous improvement in *ab initio* accuracies of spectral line intensities[12,13], which opens new possibilities for developing gas mixture and pressure standards related to accurate measurement of line intensity.

Many of these applications use cavity ring-down spectroscopy[14] (CRDS) to quantitatively measure trace and weakly absorbing species in the gas phase. Traditional CRDS systems, widely used due to their simplicity, reliability, and calibration-free nature, with inherently high sensitivity and spectral resolution, have been improved with laser and cavity stabilization technologies[15,16], and optical frequency combs providing accurate absolute frequency axes[17]. Although the best obtained relative precision exceeds $10^{-5}$, the determined line intensities can differ by up to several percent between spectrometers due to the limited ability to measure the undistorted ring-down signals[5,18]. As long as the light is turned off quickly enough, the main factors that limit the accuracy of CRDS are the nonlinearity of the amplitude and bandwidth of the detection system, which includes the detector itself, the hardware digitizing ring-downs, and electronic devices along the way. It was recently shown that digitizer nonlinearity can be reduced by the calibration to a metrology-grade reference digitizer[5]. Alternatively, one can measure the dispersion spectrum of the sample obtained from central frequencies of optical cavity modes, the positions of which are shifted within the molecular resonance range[19,20]. This calibration-free cavity mode dispersion spectroscopy (CMDS) uses only the DC part of the detection band, making it completely insensitive to the nonlinearity of the entire band. Moreover, CMDS is much less sensitive to the amplitude nonlinearity of the detector than traditional CRDS. However, due to the need for point-by-point scanning of each cavity mode profile, the CMDS is much slower than CRDS, which makes it more susceptible to various drifts. We note, however, that the CMDS speed problem can be solved, at the cost of much lower laser-to-cavity coupling efficiency due to the frequency mismatch, by using cavity buildup dispersion spectroscopy (CBDS)[21], the accuracy of which is similar to CMDS. Both approaches, CRDS and CMDS, discussed above, have recently shown the best results for the line intensity: sub-‰ accuracy in measurements[12,20] and sub-‰ agreement with *ab initio* results[12].

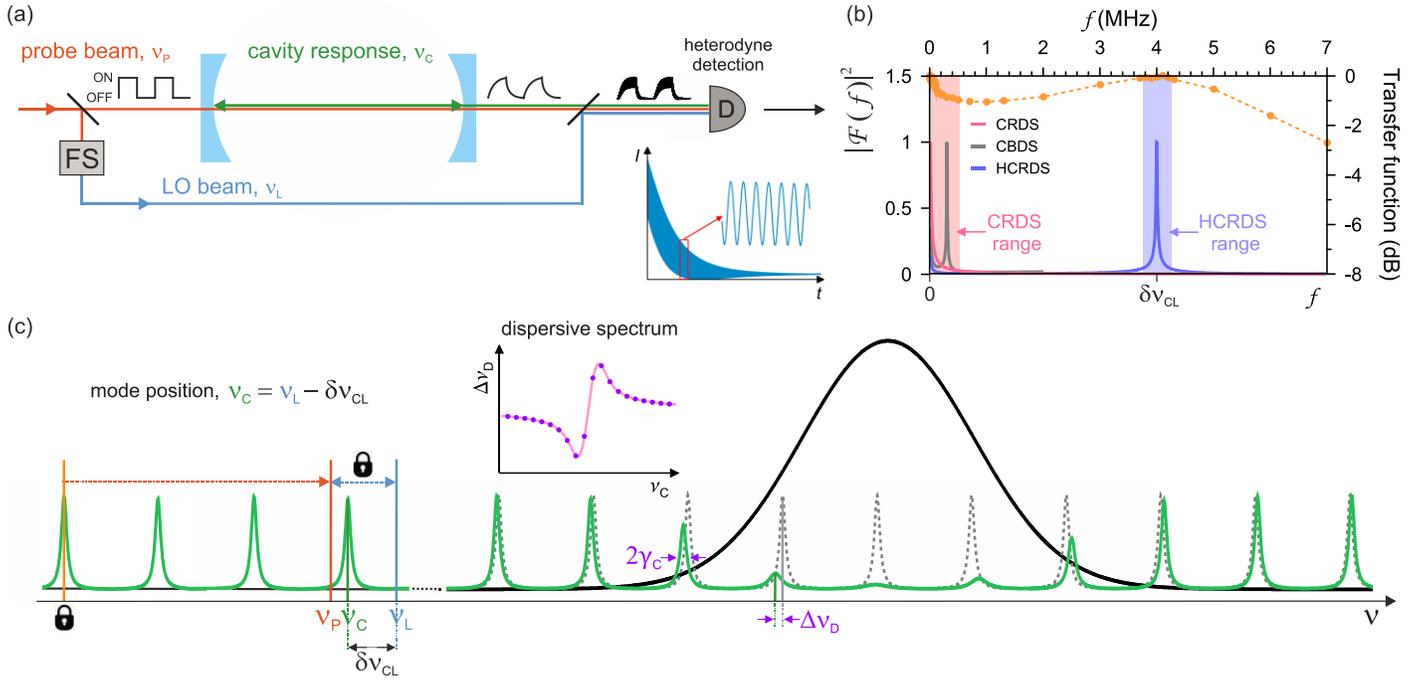

**Figure 1 | The principles of HCRDS. a**, The probe beam excites the optical cavity at arbitrary frequency $\nu_P$ close to the cavity resonant frequency, $\nu_C$. The cavity responds at the resonant frequency, $\nu_C$. The LO beam at frequency $\nu_L$ is frequency-shifted (FS) relative to the probe beam. After turning off the probe beam, it enables heterodyne detection of ring-down signals with the frequency $\delta\nu_{CL} = \nu_L - \nu_C$. **b**, The power spectrum $|\mathcal{F}(f)|^2$ of the signals obtained by CRDS, CBDS and HCRDS methods. The dots and the dashed line are the measured and fitted shape of the transfer function of the detection system used in CRDS and HCRDS measurements, respectively. The rectangles indicate the range of harmonic components of $|\mathcal{F}(f)|^2$ taken into account in the analysis of the CRDS and HCRDS signals. Fitting the Lorentz profile to the peak at $\delta\nu_{CL}$ provides the width and center of the cavity mode. **c**, In the vicinity of the molecular resonance the cavity modes are shifted by dispersion and broadened by absorption. The probe beam frequency, $\nu_P$, is tuned with steps of FSR by the broadband EOM relative to the locking point while the frequency difference $\nu_L - \nu_P$ is kept constant. For each spectral step $\nu_C$ is determined from the measured $\delta\nu_{CL}$. The frequency shifts $\Delta\nu_D(\nu_C)$ between cavity modes disturbed (continuous line) and undisturbed (dashed line) by the molecular line provide the HCRDS dispersion spectrum shown in the inset plot.

In this work, we exploit the optical frequency information carried by the ring-down cavity electromagnetic field, not explored in conventional CRDS, for high-fidelity spectroscopic measurements. Instead of measuring only the decaying light intensity, we perform heterodyne detection of ring-downs followed by Fourier analysis not only to reduce noise on the ring-down signals[22-24], but mainly to provide exact frequencies of optical cavity modes and a dispersive spectrum of a gas sample from them. We demonstrate that our approach is insensitive to light intensity measurement inaccuracies that constitute a problem for most spectrometers using light intensity detection. Moreover, it allows the selection of the most linear range of a detector transfer function, thus eliminating the major contribution to the measurement error in the traditional CRDS. We point out that with a small change in configuration, any CRDS system using laser-cavity locking technology can be easily converted into a dispersive CRDS system, providing high accuracy. In other words, dispersive CRDS combines the accuracy of CMDS[20] with the speed[25] and simplicity of conventional CRDS[14]. Using the CO line intensity as an example, we demonstrate the sub-‰ accuracy of our method, confirmed by the best *ab initio* result[12], and the long-term repeatability of our dispersion measurements at $10^{-4}$ level. For the first time, permille accuracy of the HD line intensity and permille consistency of the HD line intensity and shape with the *ab initio* results are achieved. The showcased high-accuracy examples offer promising insights into the potential application of our method in atmospheric studies[2,26], isotope ratio metrology[4,27], and the establishment of primary gas standards[8] and thermometry[28].

## Results

### The principle of heterodyne frequency detection of light decaying from an optical cavity mode

Immediate injection of probe light at the frequency $\nu_P$ into a high-finesse optical cavity begins the process of building a field inside the cavity. This always occurs at the local resonant frequency, $\nu_C$, of the cavity[29], regardless of the imperfect matching of the laser and cavity own frequency. The conventional CRDS detection system is insensitive to frequency measurement, which results in the incorrect assignment of the determined ring-down time constant, $\tau$, to the laser frequency rather than the cavity resonant frequency. Detection of the light decay relative to a stable, local oscillator (LO) beam with frequency $\nu_L$ (Fig. 1a) allows one to extract missing information about the cavity resonance frequency and guarantees the correct

frequency axis of the spectrum. The intensity emerging from the cavity is

$$I_{\text{out}}(t) = I_L + I_C e^{-t/\tau} + 2I_B e^{-t/2\tau}\cos(2\pi\, \delta\nu_{\text{CL}} t), \quad (1)$$

where the first term is the LO intensity, the second is the exponential decay of the light from the cavity measured by conventional CRDS, and the third is the heterodyne beat, with frequency $\delta\nu_{\text{CL}}$, between the LO and cavity response fields. Further analysis of this signal, in the traditional sense of heterodyne detection, assumes using a band-pass filter to reduce low-frequency technical noise. Similar results provide an analysis of the high-frequency range of the power spectrum (PS) of the signal $I_{\text{out}}(t)$, shown in Fig. 1b (see Methods). Additionally, this approach mitigates the potential heterodyne signal distortion that may arise in certain cases using electrical filters. The Lorentzian peak at frequency $\delta\nu_{\text{CL}}$ provides the position of the cavity mode $\nu_C = \nu_L - \delta\nu_{\text{CL}}$. Moreover, its full width gives the half-width of the cavity mode, $\gamma_C = (4\pi\tau)^{-1}$. High precision of measurements of both quantities is guaranteed by the high stability of the LO frequency $\nu_L$ relative to the cavity resonances. We note that the Lorentzian peak at DC frequency does not provide additional information for heterodyne cavity ring-down spectroscopy (HCRDS) presented here. More importantly, this low-frequency range of the PS, used by traditional CRDS, is usually affected by nonlinearities in the detection band. Also, the PS of the build-up signal used by CBDS[21] may encounter the same problem. Hence, in CBDS, a compromise must be achieved between detuning the laser away from the cavity mode center towards higher beat frequencies and the resulting reduction in the beam power transmitted through the cavity. The HCRDS is insensitive to the detection nonlinearity problems. It allows one to select the optimal $\delta\nu_{\text{CL}}$ frequency so that the measurement of cavity mode parameters coincides with the most linear range of the detector's bandwidth. Moreover, the symmetry of the Lorentz peak ensures that the determined cavity mode position is highly immune to nonlinearities in the light intensity measurement.

**Measurement of spectra using heterodyne cavity ring-down spectroscopy**

The idea of obtaining dispersion and absorption spectra using the HCRDS method is presented in Fig. 1c. Precise measurement of the position and width of cavity modes requires tight locking of the continuous-wave laser to the cavity. Additionally, to prevent thermal drift of the cavity modes comb, the cavity length is actively stabilized to another laser having long-term stability. The probe laser is split into two beams: one for exciting the cavity mode and the other, LO, serving as a reference for heterodyne detection of light decays. In the implemented approach, both beams are frequency-stepped using a broadband electro-optic modulator (B-EOM) and a microwave driver[23,28]. Although such a configuration generates a series of sidebands on the laser, the optical cavity acts as a spectral filter, allowing only one of the excitation beam sidebands to resonate with the cavity. Similarly, in the case of heterodyne detection, the limited detection system bandwidth allows the decay beat signal to be observed with only one sideband of the reference beam. The ring-down decays are initiated after the excitation beam is turned off by an acousto-optic modulator (AOM) (see Fig. S1a in the supplementary material). This AOM shifts the probe frequency by almost one cavity's free spectral range (FSR=204.35 MHz) and beyond the cavity resonance to avoid its influence on the ring-down signals measurement and locking the laser to the cavity. The other AOM (see Fig. S1a) shifts the LO beam to set its detuning from the probe, $\nu_L - \nu_P$, constant through the measurement of the entire spectrum. Note that laser tuning in HCRDS systems can also be realized without B-EOM, by relocking the laser to subsequent cavity modes, at the cost of lower tuning speed. This approach would result in absorptive and differential dispersive spectrum[19].

To scan the molecular spectrum, the B-EOM modulation frequency is stepped in increments equal to the FSR. Our maximum scanning range when using the first-order sideband is ±20 GHz and can be further multiplied as the sideband order increases. For each frequency step, the single heterodyne ring-down signal is acquired. Frequency scanning through the molecular spectrum is repeated. For each spectrum frequency corresponding to the center of the cavity mode, the power spectra, not the decays themselves, are averaged due to slow phase changes in the collected heterodyne light decays for that frequency. From this information, the positions and widths of the cavity modes are retrieved. Further technical details are provided in Supplementary Section S1.

The cavity mode widths provide the HCRDS absorptive spectrum with absorption coefficient $\alpha(\nu_C) = 4\pi\,\Delta\gamma_C(\nu_C)\,c^{-1}$, where $\Delta\gamma_C = \gamma_C - \gamma_{C,0} = (4\pi)^{-1} cA\,\text{Re}\{\mathcal{L}(\nu_C)\}$, $c$ is the speed of light, $\gamma_{C,0}$ is the cavity mode half-width in the absence of molecular absorption, $A$ is an area under the spectral line, and $\mathcal{L}(\nu)$ is the normalized complex-valued line-shape function so that $\int \mathcal{L}(\nu)\mathrm{d}\nu = 1$. The HCRDS dispersive spectrum is obtained from the frequency shift $\Delta\nu_D(\nu_C)$ between cavity mode positions disturbed and undisturbed by the presence of the molecular resonances. Undisturbed mode positions are fitted as a background of the dispersive spectrum. The dispersive spectrum is given by $\Delta\nu_D(\nu_C) = (4\pi n_0)^{-1} cA\,\text{Im}\{\mathcal{L}(\nu_C)\}$, where $n_0$ is a broadband refractive index of the sample. Note that the dispersive cavity mode shift $\Delta\nu_D$ and absorptive cavity mode width $\Delta\gamma_C$ are related by the Kramers-Krönig[30] relation, which yields $\Delta\nu_D \Delta\gamma_C^{-1} = n_0^{-1}\,\text{Im}\{\mathcal{L}(\nu_C)\}/\text{Re}\{\mathcal{L}(\nu_C)\}$. Accurate measurement of the line area $A$ at a known absorber concentration $N_a$ allows one to determine the line intensity $S = A/N_a$.

**HCRDS accuracy tests at the sub-permille level on CO line example**

Examples of HCRDS absorption and dispersion spectra of CO transition are shown in Fig. 2a,b. As a benchmark transition we chose one of the most accurately known molecular lines, R(23) from the 3-0 band, the line intensity of which was measured using several techniques with sub-permille accuracy[12]. Moreover, because CO is a simple diatomic molecule, its line intensity can be calculated from first principles with high accuracy[12]. Fitting the spectra with the quadratic speed-dependent hard-collision profile (qSDHCP)[31-33], results in the best agreement with the experimental line shape. The lowest standard deviation of the fit residuals achieved for dispersion was 0.14‰, and for absorption was 0.58‰. It should be noted that the qSDHCP is a variant of the currently recommended line-shape model for atmospheric data analysis[34], and the demonstrated accuracy of the laboratory dispersion spectrum is an order of magnitude greater than that required for atmospheric studies. In Fig. 2c, CO line intensities recorded using HCRDS are compared with results provided by other techniques implemented here in parallel with HCRDS, as well as with literature data and the *ab initio* calculation. A comparison of HCRDS and CMDS dispersion techniques shows an excellent 0.04‰ agreement with their averaged value. The CO R(23) line intensity, $8.0603(70) \times 10^{-25}$ cm molecule$^{-1}$, reported here has a sub-‰ relative combined uncertainty. Moreover, a comparison with CMDS data[12] from three years ago demonstrates the long-term repeatability of our dispersion measurements at $2 \times 10^{-4}$ level. Absorption measurements using HCRDS and cavity mode-width spectroscopy (CMWS)[35] provide combined uncertainties similar to dispersive methods. They introduce a small bias in line intensity but within the range of combined uncertainty. This bias is expected because absorption measurements are susceptible to nonlinearity in light-intensity detection. Interestingly, this susceptibility is two orders of magnitude lower for the dispersion techniques (see Methods). Measurement of CO transition by CRDS introduced a large bias, almost 1%, in the line intensity. Furthermore, replacing the optical detector in the detection system produced a different result. Research on the dependence of the line intensity on various configurations of the CRDS detection system has recently been carried out[5]. The main reason for the discrepant results obtained with CRDS is the nonlinearity of the detection bandwidth. We found that choosing a linear detector response range in HCRDS can reduce this nonlinearity-related error by over seven orders of magnitude (see Methods). Except for CRDS, all of our CO line intensity results agree with each other in the sub-‰ range. Moreover, our dispersion results agree up to 0.2‰ with the best CO experimental data published to date[12] and up to -0.49‰ with the best *ab initio* calculation of the CO line intensity[12].

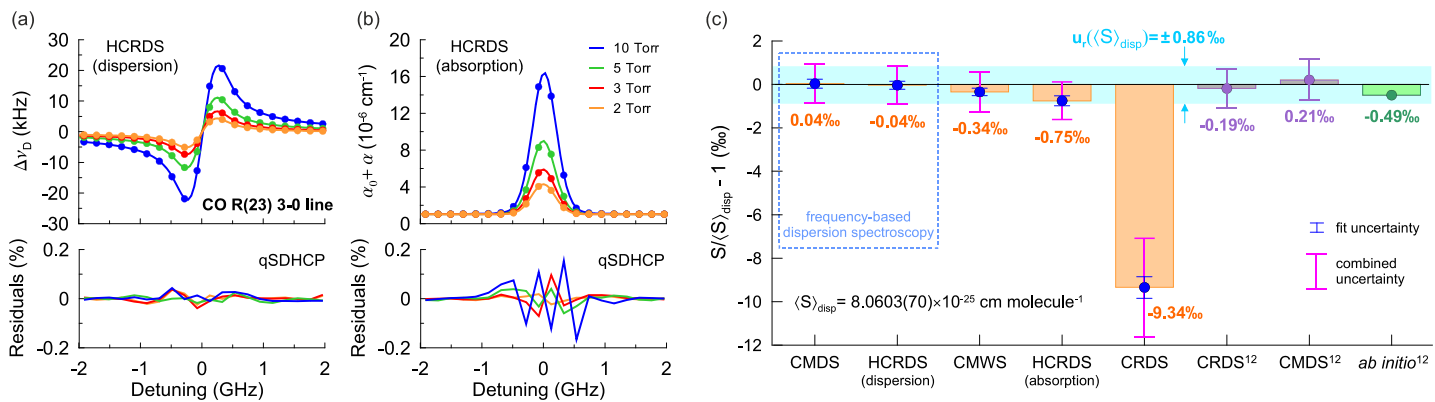

**Figure 2 | The accuracy of HCRDS. a,b,** Examples of dispersive (**a**) and absorptive (**b**) HCRDS spectra. Residuals below correspond to the difference between experimental and fitted qSDHCP line-shape model. The R(23) 3-0 line of CO ($\tilde{v}_0 = 6410.879558$ cm$^{-1}$) was measured at temperature 296 K at pressures ranging from 2 to 10 Torr. Each of the spectra presented is an average of 5000 spectral scans, and the signal-to-noise ratio ranges from 1800:1 to 5000:1. Residuals are presented relative to the line profile amplitude at 10 Torr. Typical measurement time for one scan is 60 ms. The offset frequency of 192193.256628 GHz is subtracted from the abscissa. **c,** Comparison of HCRDS results for the intensity of CO R(23) 3-0 line with CMDS, CMWS and CRDS measurement results in this work as well as literature data[12] and *ab initio* calculation[12]. The reference line intensity $\langle S \rangle_{disp}$ is the average of the dispersion measurement results with CMDS and HCRDS. These dispersive measurements provide the relative combined uncertainty of 0.86‰. Combined uncertainty includes both type A (fit uncertainty) and type B uncertainties. The large deviation seen for CRDS result is caused by the nonlinearity of the detection bandwidth. The remaining experimental results are in sub-‰ agreement with each other as well as with the best reported data and the *ab initio* results.

**Permille-level accuracy spectroscopy of HD line shape**

Dispersive CRDS spectroscopy reveals its potential for the fundamental studies of collisions between hydrogen molecules through accurate measurements of the shape of its spectral lines. In Fig. 3a, we show the HCRDS dispersion spectrum of the P(3) 2-0 HD transition. The qSDHCP does not fully describe the experimental spectrum for this system, which is revealed as a systematic structure on the fit residuals. For lighter molecules such as HD, attention needs to be paid to the proper description of the velocity-changing collisions leading to Dicke narrowing of the line. The speed-dependent billiard-ball profile (SDBBP)[36] provides a more physical description of molecular collisions and allows the HD spectrum to be modeled down to random noise in residuals with a standard deviation of 0.6‰. Calculations of SDBBP at low pressures used here require the implementation of an iterative approach[37]. We note that the selected HD transition, unlike the CO transition, is weaker and not so well isolated from other transitions (Fig. 3b). Our results for the HD line intensity and their comparison with the theory are shown in Fig. 3c. The reference is the line intensity of $3.1793(40) \times 10^{-26}$ cm molecule$^{-1}$ obtained from the SDBBP analysis, with a relative combined uncertainty of 1.3‰, in which the line-shape parameters were fixed to *ab initio* values calculated in this work (see Supplementary Section S2). If the line-shape parameters were fitted, SDBBP gave a small deviation of 2.55‰ from the reference line intensity. The results for the qSDHCP show a systematic bias of up to -5.23‰ compared to the reference value due to the incorrect model used to analyze the experimental spectrum. The HD line intensity reported here is one of the most accurate experimental results to date[38,39]. Moreover, its comparison with *ab initio* calculations of line intensity provided by Ref.[40] and Ref.[41] gives excellent agreement up to -0.73‰ and 1.79‰, respectively. HCRDS demonstrates the highest accuracy in the experimental study of the molecular shapes of hydrogen lines. For the first time, spectroscopy confirmed both the *ab initio* intensity and the *ab initio* shape of the molecular hydrogen line, reaching permille accuracy.

## Discussion

The high accuracy of dispersive CRDS makes it an ideal tool for many existing, demanding applications of ultrasensitive optical spectroscopy, where an accurate measurement of the shape and intensity of the molecular line is critical. A very insidious systematic error of the traditional CRDS method may come from the distortion of the detection band that modifies the decay time constant but not the exponential shape of the decay. A recently proposed solution for CRDS is to calibrate its detection system[5]. Although this ultimately provides accuracy similar to the dispersive CRDS presented here, the calibration process can only be performed in a few metrology institutes worldwide. Our dispersive CRDS technique is calibration-free and, in principle, does not require complex modifications to existing frequency-stabilized CRDS systems. More importantly, with the appropriate selection of the linear range of the detection band, this technique is practically insensitive to detection band nonlinearities. Also, its susceptibility to signal amplitude nonlinearity is two orders of magnitude smaller than in traditional CRDS. Moreover, the HCRDS dispersion spectrum is completely measured in frequencies, which allows it to be referenced to atomic frequency standard. This approach will ensure SI-traceability and can significantly facilitate the comparison of inter-laboratory data.

Analyzing the uncertainty budget of the determined line intensities (see Methods), it can be seen that the main contribution comes from pressure measurement and the uncertainty of the sample composition. This provides the unique opportunity to use the dispersive CRDS technique to develop new optical standards for gas composition and pressure based on the measurement of the spectral line intensity. Gas pressure measurement based on the resonant refractive index (i.e., spectroscopic measurement), rather than non-resonant as in current cavity-based refractometry, has the advantage of being much less sensitive to cavity deformation and diffraction effects (Gouy phase shift)[8]. These factors affect the shift of the entire comb of cavity modes, in addition to the pressure-induced changes. Although refractometric measurements of cavity mode positions relative to their vacuum positions pose a challenge, this issue does not affect the measurement of local shifts in cavity modes near molecular resonances. We expect that further development of *ab initio* line intensity calculations will drive progress in this field. A significant contribution to the uncertainty budget of line intensity also comes from choosing the correct line-shape model. Accurate dispersive CRDS spectra can help in testing new theoretical models of intermolecular interactions and *ab initio* calculations of line intensities and line-shape parameters. As we have shown, the qSDHCP profile is no longer sufficient to accurately describe the shape of molecular hydrogen lines, and the more physical SDBBP should be used instead. The intensity of the HD line obtained using this model, combining *ab initio* line-shape parameters, agrees excellently with *ab initio*-calculated line intensity. Moreover, the results from Fig. 2c clearly show that agreement between the current most accurate experimental results for CO line intensity is better than their agreement with the most accurate *ab initio* result. Therefore, line intensities determined by our method can serve as a sub-‰ reference used in such important applications as global GHG measurements, studies of the atmosphere of exoplanets or monitoring trace humidity in semiconductor production. Finally, the presented dispersive spectroscopy stimulates an area of research devoted to very accurate *ab initio* calculations of spectral line shapes and their intensities, which have implications for both fundamental science and many applications.

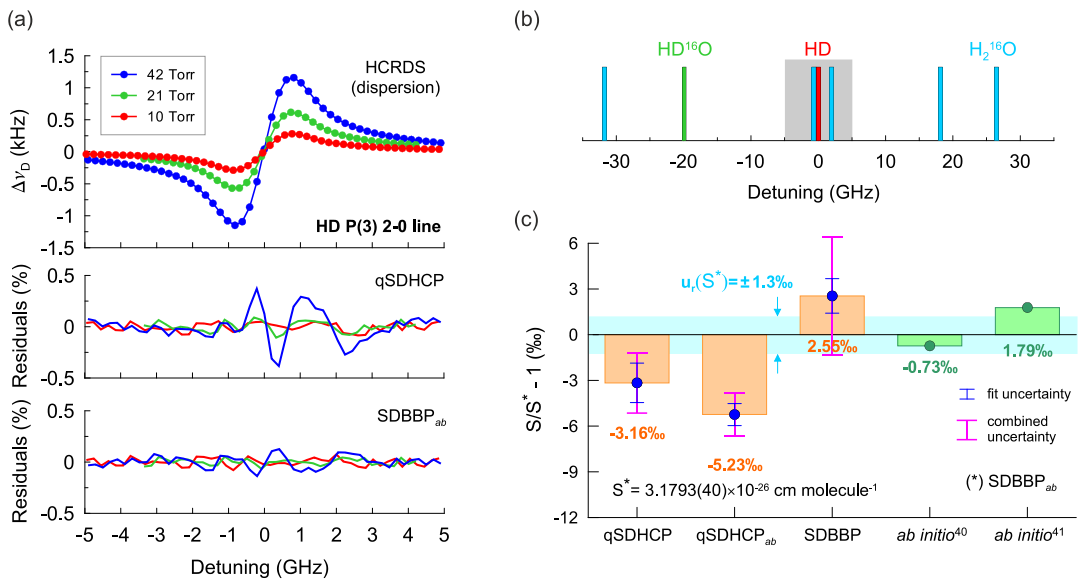

**Figure 3 | Permille-level accuracy spectroscopy of the molecular hydrogen line shape. a,** The dispersive HCRDS spectrum of the P(3) 2-0 line of HD ($\tilde{\nu}_0 = 6798.7679$ cm$^{-1}$) measured at temperature 296 K at pressures ranging from 10 to 42 Torr. The residuals below correspond to the difference between the experimental model and the fitted line-shape model. In fits qSDHCP and SDBBP$_{ab}$ were used. The subscript $ab$ indicates that line-shape parameters such as collisional width and shift, Dicke narrowing and speed dependence of collisional width and shift were set to the *ab initio* values. Each spectrum is an average of 5000-15000 spectral scans, and the signal-to-noise ratio ranges from 700:1 to 2000:1. Residuals are presented relative to the line profile amplitude at 42 Torr. Typical measurement time for one scan is 750 ms. The offset frequency of 203821.934011250 GHz is subtracted from the abscissa. **b,** Transitions in H$_2^{16}$O and HD$^{16}$O molecules that were included in the analysis of the HD P(3) 2-0 line. The rectangle indicates the spectral range of the HD line measurement. **c,** Line intensity measurement results of the HD spectrum from (**a**) and their comparison with the *ab initio* results[40,41]. The reference line intensity $S^*$ corresponds to SDBBP$_{ab}$ analysis with a relative combined uncertainty of 1.23‰. Combined uncertainty includes both type A (fit uncertainty) and type B uncertainties. Results of analysis using qSDHCP, qSDHCP$_{ab}$ and SDBBP are also shown. Permille agreement with the *ab initio* intensity and shape of the HD line is presented.

## Methods

### The influence of light detection nonlinearity on the accuracy of spectral line intensity measurement

In order to gain a comprehensive understanding of how distortions in the ring-down decay signals, induced by the nonlinearity of light intensity detection, impact the accuracy of determined spectral line intensities, it was neccesary to initially model the temporal response of the cavity to laser beam deactivation. Subsequently, we simulated conditions close to the experimental ones, in which nonlinear light detection effects modify the measured intensity of the light, which emerges from the cavity.

Immediately injecting light at the frequency $\omega$ into the cavity results in a cavity response at the frequency $\omega_C$ with temporal inertia described by $\Gamma_C^{-1}$. The electric field leaving the cavity is a superposition of the laser and cavity fields:

$$E_{\text{out}}^{\text{ON}}(t) = E_{\text{out}}^0 \frac{\Gamma_C^0}{\Gamma_C - i\delta\omega}\left(e^{-i\omega t} - e^{-\Gamma_C t}e^{-i\omega_C t}\right), \quad (2)$$

where $\delta\omega = \omega - \omega_C$ is angular frequency detuning of the laser beam from the cavity mode center, the angular frequency $\Gamma_C = -t_r^{-1}\ln(Re^{-\alpha L})$ relates to the measured cavity mode half width at half maximum $\gamma_C = (2\pi)^{-1}\Gamma_C$, $R$ is reflectance of the cavity mirror, $\alpha$ is the absorption coefficient of the intra-cavity gas sample, $\Gamma_C^0 \approx t_r^{-1}(1-R)$ is $\Gamma_C$ for $\alpha = 0$ and $t_r^{-1} = $ FSR. Note that the complex Lorentz function $(\Gamma_C - i\delta\omega)^{-1}$, characterized by $\alpha$-dependent width $\Gamma_C$, relates to the cavity mode shape and quantizes the fraction of the laser field which is transmitted through the cavity when the laser frequency is detuned by $\delta\omega$ away from the cavity mode center. The structure of Eq. (2) indicates the interference pattern with the modulation frequency $\delta\omega$ that will appear on the light transmitted through the cavity in typical cavity-enhanced systems before a steady state is reached[21].

When the laser beam is turned off at a time $t_0$, the process of the intracavity light decay begins. Assuming that the laser field is turned off at the rate of $\Gamma_0$, the field passing through the cavity has a two-part form:

$$E_{\text{out}}^{\text{OFF}}(t) =$$
$$E_{\text{out}}^0 e^{-i\omega t_0}\Gamma_C^0\left\{\frac{1 - (1 - \Gamma_C/\Gamma_0 + i\,\delta\omega/\Gamma_0)e^{-(\Gamma_C - i\delta\omega)t_0}}{(1 - \Gamma_C/\Gamma_0 + i\,\delta\omega/\Gamma_0)(\Gamma_C - i\delta\omega)} \times\right.$$
$$\left. e^{-(\Gamma_C + i\omega_C)(t-t_0)} + \frac{1}{\Gamma_C - \Gamma_0 - i\delta\omega}e^{-(\Gamma_0 + i\omega)(t-t_0)}\right\}, \quad (3)$$

where the first part describes the exponential decay of the cavity mode field with the time-constant $\Gamma_C^{-1}$ and the second part is the adopted model of the exponential decay of the laser field. The amplitude of the laser field passing through the cavity depends on the detuning $\delta\omega$ of the laser from the center of the cavity mode, the profile of which is defined as before by the complex Lorentz function, but with a width modified to $\Gamma_C - \Gamma_0$. Note that even if initially the laser and cavity are not frequency matched, the ring-down decay is always observed at the cavity mode frequency $\omega_C$. This observation, as well as Eq. (3), are in good agreement with the calculations provided in Ref.[42]. If the laser light is turned off very quickly such that $\Gamma_0 \gg \Gamma_C$ and $\Gamma_0 \gg \delta\omega$, the first term in Eq. (3) dominates and the expression for $E_{\text{out}}^{\text{OFF}}(t)$ can be reduced to a simple single-exponential decay form

$$E_{\text{out}}^{\text{OFF}}(t) = E_{\text{out}}^{\text{ON}}(t_0)e^{-(\Gamma_C + i\omega_C)(t-t_0)}. \quad (4)$$

However, if the laser turn-off rate $\Gamma_0$ is relatively small and/or the extinction ratio of the laser field amplitude is low, e.g. 50 dB[43], then Eq. (3) clearly shows that the recorded light decay will be distorted as a result of interference at the frequency $\delta\omega$ between the laser and cavity fields. In our CRDS and HCRDS measurements, this situation did not occur because $\Gamma_0$ was more than 600 times larger than $\Gamma_C$ and the extinction ratio of the laser field was ~80 dB.

In the HCRDS method, the ring-down decay field described by Eq. (4) is beaten with the laser field of the local oscillator (LO). The field incident on the detector is

$$E_{\text{out}}^{\text{OFF}}(t) = E_{\text{out}}^{\text{ON}}(t_0)e^{-(\Gamma_C + i\omega_C)(t-t_0)} + E_L e^{i\varphi_L}e^{i\omega_L(t-t_0)}, \quad (5)$$

where $E_L$, $\omega_L$ and $\varphi_L$ are the amplitude, angular frequency, and phase of the LO field. The corresponding light intensity $\sim |E_{\text{out}}^{\text{OFF}}|^2$ is

$$I_{\text{out}}(t) = I_L + I_C e^{-2\Gamma_C t} + 2I_B e^{-\Gamma_C t}\cos(\delta\omega_{\text{CL}} t), \quad (6)$$

where $\delta\omega_{\text{CL}} = \omega_C - \omega_L$ is the beat frequency between the cavity and LO fields. If $\omega_L$ is known, the cavity mode position can be easily calculated from the measured value of $\delta\omega_{\text{CL}}$. We note that $\omega_L$ needs to be known only on a local frequency axis associated with the cavity modes. Analysis of the Lorentz peak at the frequency $\delta\omega_{\text{CL}}$ in the power spectrum of the light intensity $I_{\text{out}}$ provides information about both the dispersion shift of the cavity mode and its absorption broadening.

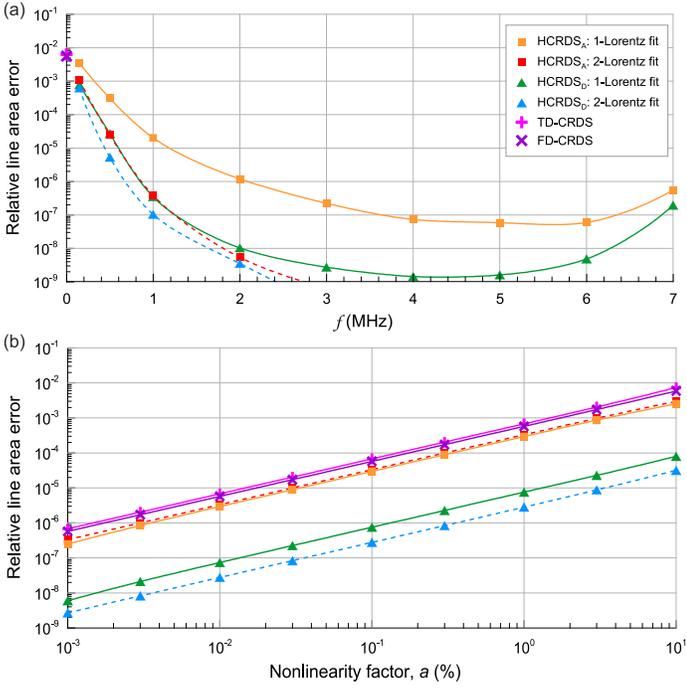

**Figure 4 | Nonlinearity of light detection. a**, The influence of the detection system transfer function, from Fig. 1b, on the accuracy of the line area determined from the simulated absorption and dispersion line profiles in HCRDS and CRDS. The $f$ axis corresponds to the beat frequencies between the LO laser and the cavity response (for CRDS $f = 0$). HCRDS results are presented for both absorption (HCRDS$_A$) and dispersion (HCRDS$_D$). Data marked as 1-Lorentz and 2-Lorentz correspond to different models used in the analysis of power spectra of heterodyne ring-downs. Results for traditional CRDS were obtained from both time (TD-CRDS) and frequency (FD-CRDS) domain analysis. **b**, The influence of the nonlinearity of the detection system amplitude on the accuracy of the line area determined from simulated absorption and dispersion line profiles in HCRDS and CRDS. Results for different values of the nonlinearity parameter, $a$, are presented.

The transfer function (TF) of the detection system of the spectrometer, described in Supplementary Section S1, was measured as its amplitude response to changes in the beat signal frequency of the detected light. We developed the TF model as a linear combination of complex-valued high-pass and low-pass filter functions. The parameters of this model were selected to reproduce best the experimental shape of the TF shown in Fig. 1b. To test the line intensity accuracy against the nonlinearity of the detection system's bandwidth, the absorption and dispersion profile was simulated with the line intensity and shape parameters corresponding to the R(23) 3-0 CO transition at a pressure of 10 Torr. The light decay signal was simulated using Eq. (6) for each frequency and Fourier transformed. The result was multiplied by our complex-valued TF model and inverse Fourier transformed to reproduce the ring-down decay having a different shape than the original exponential one. The power spectra of the reproduced heterodyne ring-down decays were then analyzed, focusing on the frequency range around the frequency $\delta\omega_{CL}$. Two models, a single Lorentz profile centered on the frequency $\delta\omega_{CL}$ and two Lorentz profiles centered on $\delta\omega_{CL}$ and zero frequencies, were included in the analysis to provide the beat frequency parameter $\delta\omega_{CL}$ (for HCRDS dispersion) and cavity mode width (for HCRDS absorption). Finally, the line area determined from the obtained absorption and dispersion profiles was compared with the simulated value, and the systematic deviation from it was calculated. The entire procedure was repeated for several beat frequencies up to 7 MHz. The results are shown in Fig. 4a. In the same figure, we also present results for CRDS. The procedure for simulating distorted light decays was similar to that described above for HCRDS, starting with single-exponential decay signals. Their analysis was performed both traditionally in the time domain, determining the time constant, and in the frequency domain using the Lorentz peak of the power spectrum on the zero frequency to determine the cavity mode width. The results for HCRDS are clearly better than those for CRDS for all frequencies in the tested range of $\delta\omega_{CL}$. Note that the greatest advantage of HCRDS over CRDS at the relative accuracy level $10^{-7}$ is visible for the most linear range of the detection band around the frequency of 4 MHz. Furthermore, dispersive HCRDS, with an accuracy level $> 10^{-9}$, appears to be more accurate than absorptive one. Using the model with simultaneous fitting of two Lorentz profiles further improves the accuracy of the retrieved line intensity, especially in the case of absorption HCRDS.

In tests of the line intensity accuracy in terms of the nonlinearity of the detection system amplitude, the heterodyne ring-down decays simulated using Eq. (6) were multiplied by the nonlinearity function[21] $y(t) = 1 - a\, I_{out}(t)/I_{out}^{max}$, where $I_{out}^{max}$ is the maximum amplitude of $I_{out}(t)$ and $a$ is the amplitude nonlinearity parameter in %. As before, the heterodyne ring-down decays were simulated at each frequency in the simulated CO spectrum. Absorption and dispersion profiles were obtained from them and the determined line area was compared with the simulated one. The results are shown in Fig. 4b. In the simulation for HCRDS we chose $\delta\omega_{CL} = 4$ MHz. To analyze the Lorentz peak in the power spectrum, as before, both the single Lorentz profile centered on the frequency $\delta\omega_{CL}$ and two Lorentz profiles centered on $\delta\omega_{CL}$ and 0 frequencies were used. The CRDS results included the decay signal analysis in both time and frequency domains. From Fig. 4b it can be seen that the method of analyzing the light decays in CRDS does not affect the dependence of the line intensity accuracy on $a$ parameter. HCRDS absorption results are similar to CRDS results. However, for dispersive HCRDS we observe a two-order improvement in line intensity accuracy compared to CRDS and absorption HCRDS. This immunity to nonlinear detection is due to the fact, that any deformations of the Lorentz peak amplitude, but symmetrical about its center, will not affect its position. As before, the 2-Lorentz fitting model further improves the line intensity accuracy.

## Uncertainty budget for CO and HD line intensities

The intensity of the spectral line can be calculated as $S = A/N_a$, where $A$ is the line area determined from the line shape analysis and $N_a$ is the concentration of absorbers (number of molecules per volume) determined as a $\kappa$ fraction of the total gas concentration $N$, $N_a = \kappa N$. Measuring the total pressure $p$ of the gas sample and its temperature $T$, the total gas concentration $N$ can be determined from the ideal gas law $N = p(k_B T)^{-1}$, where $k_B$ is the Boltzmann constant. As a result we obtain the following formula for the line intensity $S$ measured at temperature $T$, $S(T) = k_B T A(\kappa p)^{-1}$. Note that the intensity of the line generally depends on the temperature. To take this into account, we define a temperature-dependent function $f_S(T) = \frac{S(T_0)}{S(T)}$, where $T_0$ is the reference temperature. The final expression for the line intensity determined for the reference temperature $T_0$ from a spectrum measured at temperature $T$ is

$$S(T_0) = k_B f_S(T) T A (\kappa p)^{-1}. \qquad (7)$$

The function $f_S(T)$ can be calculated based on the ratio of total internal partition functions provided by the HITRAN database[44].

To estimate the combined uncertainty $u(S)$ of the line intensity determined for $T_0$, it should be noted that the quantities $f_S(T)$ and $T$ in Eq. (7) are mutually dependent. Hence, the non-zero covariance of these quantities $C_{T, f_s} = \left(\frac{\partial f_s}{\partial T}\right) u^2(T)$ must be taken into account in the calculation of $u(S)$. Assuming that the remaining quantities in Eq. (7) are independent of each other, we found the following formula for

**Table 1 | Uncertainty budget for CO line intensity.** This table contains the quantities $x$ taken into account when estimating the combined CO line intensity uncertainty using equation (8). Type A standard uncertainties and type B standardized uncertainties of $x$, and the corresponding combined uncertainties $u(x)$ are given. Uncertainties of the line area, $A$, are shown for used measurement methods. The combined uncertainty of the line intensity $u(S)$ is given for each method. We also show the value of $u(\langle S \rangle)$ estimated for the mean line intensity $\langle S \rangle$ obtained from the CMDS and HCRDS dispersion measurement results. For each measurement method, the bias of the line intensity from $\langle S \rangle$ is also shown. All values are given in promilles.

| Method | Quantity, $x$ | Type A (‰) | Type B (‰) | $u(x)/x$ (‰) | $u(S)/S$ (‰) | $u(\langle S \rangle)/\langle S \rangle$ (‰) | Bias from $\langle S \rangle$ (‰) |
|---|---|---|---|---|---|---|---|
| | T | < 0.01 | 0.1 | 0.1 | | | |
| | p | 0.05 | 0.62 | 0.62 | | | |
| | κ | | 0.44 | 0.44 | | | |
| CMDS | A | 0.20 | 0.32 | 0.38 | 0.91 | 0.86 | 0.04 |
| HCRDS (dispersion) | A | 0.18 | 0.24 | 0.30 | 0.88 | | -0.04 |
| CMWS | A | 0.17 | 0.39 | 0.43 | 0.93 | | -0.34 |
| HCRDS (absorption) | A | 0.23 | 0.11 | 0.26 | 0.87 | | -0.75 |
| CRDS | A | 0.50 | 2.04 | 2.10 | 2.30 | | -9.34 |

**Table 2 | Uncertainty budget for HD line intensity.** This table contains the quantities $x$ taken into account when estimating the combined HD line intensity uncertainty using equation (8). Type A standard uncertainties and type B standardized uncertainties of $x$, and the corresponding combined uncertainties $u(x)$ are given. Uncertainties of the line area $A$ are shown for the various profiles we used in line shape analysis. The combined uncertainty of the line intensity $u(S)$ is given for each profile. For each profile, the bias of the line intensity from $S^*$, obtained from SDBBP analysis with *ab initio* values of line-shape parameters, is also shown. All values are given in promilles.

| Profile | Quantity, $x$ | Type A (‰) | Type B (‰) | $u(x)/x$ (‰) | $u(S)/S$ (‰) | Bias from $S^*$ (‰) |
|---|---|---|---|---|---|---|
| | T | < 0.01 | 0.1 | 0.1 | | |
| | p | 0.05 | 0.62 | 0.62 | | |
| | κ | | 1 | 1 | | |
| SDNGP | A | 1.29 | 0.93 | 1.59 | 1.98 | -3.16 |
| SDNGP (ab initio) | A | 0.72 | < 0.01 | 0.72 | 1.38 | -5.23 |
| SDBBP | A | 1.13 | 3.51 | 3.69 | 3.87 | 2.54 |
| SDBBP * (ab initio) | A | 0.36 | < 0.01 | 0.36 | 1.23 | |

the square of the relative uncertainty of the line intensity determined for the temperature $T_0$

$$\frac{u^2(S)}{S^2}\Big|_{T_0} = \left[\frac{1}{T^2} + \frac{1}{f_s^2}\left(\frac{\partial f_s}{\partial T}\right)^2 + 2\frac{1}{Tf_s}\frac{\partial f_s}{\partial T}\right]u^2(T) + \frac{u^2(p)}{p^2} + \frac{u^2(A)}{A^2} + \frac{u^2(\kappa)}{\kappa^2}. \quad (8)$$

This expression was used to estimate the relative combined uncertainties of CO and HD line intensities reported here. The corresponding line intensity uncertainty budgets are presented in Tables 1 and 2. It should be noted that when estimating the combined uncertainty of the CO line intensity, which is the average of the results of the dispersion HCRDS and CMDS methods, only the measurements of $A$ could be treated as independent, while the measurement of $T, p$, and $\kappa$ was common to both methods. As a result, the combined uncertainty of the average line intensity $u(\langle S \rangle)$ was calculated by inserting the relative standard deviation $u(A)/A$ of the mean value of $A$ from both measurement methods into Eq. (8). Let us also note that in the case of HD molecule, the quantity $\kappa$ is determined not by the isotope composition as it is for CO, but by the purity of the HD sample.

It is worth mentioning how deviations from the ideal gas law affect the line intensity values reported here. We limit ourselves to the first correction in the expanded gas law $p = Nk_BT(1 + B(T)N + C(T)N^2 + \cdots)$, where $B(T)$, $C(T)$ are virial coefficients of the second and third order, which reduces the expression for $N$ to the form $N = p(k_BT)^{-1}(1 - pp_n^{-1}T_nT^{-1}V_n^{-1}B(T))$, where $p_n$, $T_n$, $V_n$ describe normal conditions. For the measured CO and HD transitions the virial coefficients $B(T)$ determined at a temperature of 296 K are $-8 \text{ cm}^3/\text{mol}$ (Ref.[45]) and $+14 \text{ cm}^3/\text{mol}$ (Ref.[46]), respectively. They correspond to CO and HD line intensity relative corrections of $-4.334 \times 10^{-6}$ and $3.034 \times 10^{-5}$, which were calculated for the highest pressures of 10 Torr and 40 Torr, at which the spectra of CO and HD molecules were measured. These changes are at least two orders of magnitude lower than the accuracy of the determined here line intensities, so they are not included in the uncertainty budget.

## Acknowledgments

A.C., H.J., D.L. were supported by the National Science Centre in Poland through Project No. 2020/39/B/ST2/00719. S.W. was supported by the National Science Centre in Poland through Projects No. 2020/39/B/ST2/00719 and 2021/42/E/ST2/00152. N.S. was supported by the National Science Centre in Poland through Project No. 2019/35/B/ST2/01118. K.B. was supported by the National Science Centre in Poland through Project No. 2018/30/E/ST2/00864. R.C. was supported by the National Science Centre in Poland through Project No. 2021/41/B/ST2/00681. For the purpose of Open Access, the author has applied a CC-BY public copyright licence to any Author Accepted Manuscript (AAM) version arising from this submission. P.W. is funded by the European Union (ERC-2022-STG,H2TRAP,101075678). Views and opinions expressed are however those of the author(s) only and do not necessarily reflect those of the European Union or the European Research Council Executive Agency. Neither the European Union nor the granting authority can be held responsible for them. The research is part of the program of the National Laboratory FAMO in Toruń, Poland.

## Author contributions

A.C. and D.L. concieved the idea of eliminating systematic errors in CRDS using heterodyne dispersive cavity ring-down spectroscopy. A.C. and D.L. designed the concept of the research and performed measurements. A.C., S.W., G.K., N.S., K.B., D.L. prepared the experimental setup. A.C., S.W., R.C., D.L. performed data analysis. H.J. and P.W. performed ab initio calulations. A.C. wrote the original draft of the manuscript, and all authors contributed to the final manuscript. A.C. and D.L coordinated the project.

## Competing interests

The authors declare no competing interests.

## Additional information

The manuscript contains supplementary material.

# Supplementary material

## Heterodyne dispersive cavity ring-down spectroscopy exploiting eigenmode frequencies for high-fidelity measurements

Agata Cygan, Szymon Wójtewicz, Hubert Jóźwiak, Grzegorz Kowzan, Nikodem Stolarczyk, Katarzyna Bielska, Piotr Wcisło, Roman Ciuryło, Daniel Lisak

*Institute of Physics, Faculty of Physics, Astronomy and Informatics, Nicolaus Copernicus University in Toruń, Grudziadzka 5, 87-100 Torun, Poland*

### Section S1. HCRDS experiment

A sketch of the experimental system implementing the HCRDS method is shown in Fig. S1a. To measure the shapes of the CO and HD lines, two external cavity diode lasers (*Toptica CTL*) were used, covering the spectral ranges of 1520-1630 nm and 1460-1570 nm, respectively. In each case, the laser beam was divided into two orthogonally polarized beams. An s-polarized beam, phase-modulated at 20 MHz by a narrow-band electro-optic modulator (EOM), was used to lock the laser to the optical cavity using the Pound-Drever-Hall scheme[1]. The frequency stability of this lock was at the Hz level. A small portion of the p-polarized laser beam was transmitted via optical fiber to an optical frequency comb system (*Menlo Systems*) to enable continuous measurement of the absolute laser frequency. The main part of the p-polarized beam was further split into two beams - one for excitation of the optical cavity mode and the other for the reference heterodyne detection of light decays from the cavity. Both beams were frequency-stepped using a broad-band electro-optic modulator (B-EOM) and a microwave driver[2]. Although such configuration generates a series of sidebands on the laser, the optical cavity acts as a spectral filter, allowing only one sideband of the excitation beam to resonate with the cavity. Similarly, in the case of heterodyne detection the limited detection system bandwidth allows the observation of the decay of the beat signal with only one sideband of the reference beam. We chose a first-order sideband that enables scanning of the excitation and reference beams in the frequency range of 0.2 – 20 GHz. Note, however, that the scanning range can be further multiplied by choosing a higher-order sideband. The ring-down decays were initiated after turning off the excitation beam by an acousto-optic modulator (AOM) driven by frequency $f_A$. This AOM also shifts the carrier frequency by almost one free spectral range of the cavity (FSR ≈ 204.35 MHz) and beyond the cavity resonance to avoid its influence on the measurement of ring-down signals and locking the laser to the cavity. Another AOM detunes the frequency of the reference beam from the frequency of the excitation beam by a constant value of $\delta\nu_{PL}$ equal to several MHz.

The optical cavity was formed by two high-reflectivity mirrors placed at a distance of ∼73 cm from each other. The mirrors are dual-wavelength coated to use a separate laser wavelength (1064 nm) to stabilize the cavity length[3]. This was done with respect to the I$_2$-stabilized Nd:YAG laser characterized by long-term frequency stability of 1 kHz. The double-pass acousto-optic modulator system[4] and phase-sensitive detection were used to generate an error signal in the cavity length locking loop servomechanism. Two sets of cavity mirrors were used for HCRDS measurements - one for the CO molecular system and the other for HD. They were characterized by a reflectance coefficient of 0.999925 and 0.999993, corresponding to full widths at half maximum of the cavity mode of 4.9 kHz and 0.43 kHz, respectively. They lead to cavity finesses of 41900 and 449000 and ring-down time constants of 32 μs and 350 μs, respectively. In R(23) 3-0 CO transition measurements, we used a commercial sample of CO with purity of 0.99997, produced by the reaction of water with petrogenic natural gas having an estimated $^{13}\delta C_{VPDB}$ isotope content of -40‰ with respect to the Vienna PeeDee Belemnite scale[5]. In P(3) 2-0 HD transition measurements, we used a commercial HD sample characterized by a purity of 97.0%. During the measurements, the sample gas pressure was measured in parallel using three manometers *(MKS Baratron 690A* with 10 and 100 Torr full range and *Mensor CPG2500* with 900 Torr full range) with the highest relative combined uncertainty of $6 \times 10^{-4}$. The cavity temperature was stabilized at 296 K with a total standard uncertainty of 30 mK. Analog ring-down decays, measured with a DC-coupled photodetector (*New Focus 2053*) with a bandwidth of up to 7 MHz, were digitized with an oscilloscope card characterized by a bandwidth of 100 MHz and 14-bit vertical resolution (*National Instruments PCI-5122*).

To scan the molecular transition the B-EOM modulation frequency was stepped in increments of the FSR. A list of necessary frequencies was prepared and loaded into the microwave generator's memory before each spectrum measurement for fast frequency switching. Then the TTL signal controlled the determination of subsequent frequencies at the generator output. The time sequences of the TTL signals used in the experiment to control the onset of ring-down signal measurements and data acquisition process and to determine the frequency switching time of the microwave driver are shown in Fig. S1b. The typical TTL signal period was ~2 ms for CO measurements and ~15 ms for HD measurements. During this time the heterodyne decay signal was acquired, the frequency of the microwave generator was changed, and the cavity was pumped with light at the new frequency. The frequency switching time of the microwave generator was <1 ms. The size of the frequency list loaded into the microwave generator's memory allowed the entire spectrum to be measured several dozen times without interruption. After collecting the appropriate amount of data, the initial averaging process began. For each frequency in the spectrum, the power spectra, not the decays themselves, were averaged due to slow phase changes in the collected heterodyne light decays for that frequency. From the finally averaged power spectra of heterodyne ring-down signals, information on the positions and widths of cavity modes was obtained, and from them, the absorption and dispersion spectra were obtained. For CO measurements, 5000 spectral scans were collected, while for HD, it was 5000-15000.

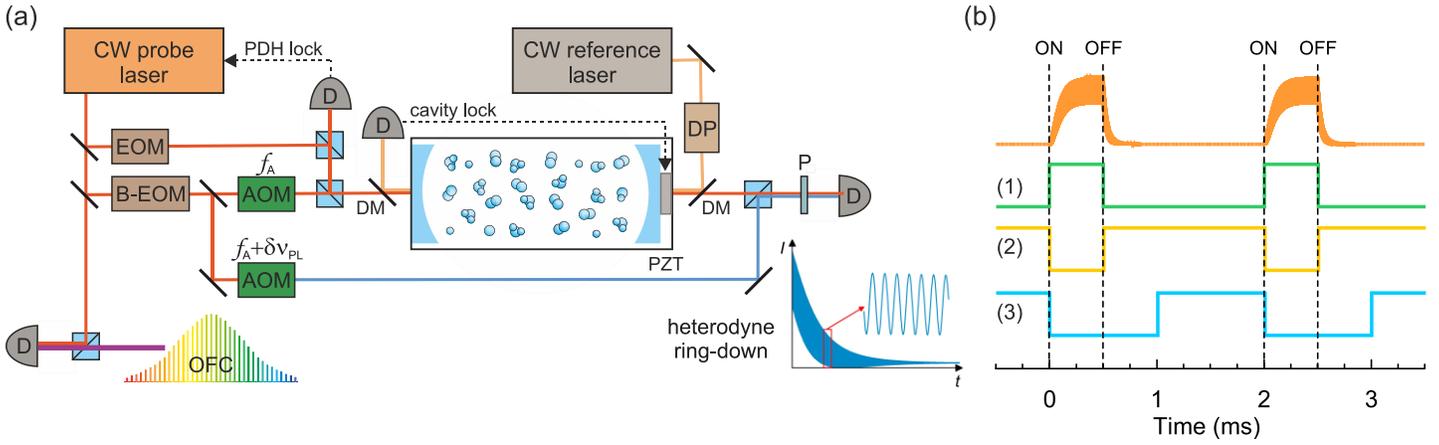

**Figure S1 | HCRDS experiment. a**, A continuous-wave (CW) probe laser is locked to the cavity by the Pound-Drever-Hall (PDH) method. An electro-optic modulator (EOM) modulates the laser light phase to generate the PDH error signal. The cavity length is stabilized with respect to another CW reference laser. The absolute frequency of the laser is measured using the optical frequency comb (OFC). The intracavity gas sample is probed using a single sideband of the broad-band electro-optic modulator (B-EOM) scanning in the microwave range. Acousto-optical modulators (AOM) are used to detune the carrier frequency from the cavity resonance by $f_A$, and also to detune the local oscillator beam from the probe beam by $\delta\nu_{PL}$. The inset presents an example of a heterodyne ring-down signal. Explenation of abbreviations: D – photodetector, DM – dichroic mirror, PZT – piezo transducer, OFC – optical frequency comb, P –polarizer, DP – double-pass AOM system. **b,** Top: heterodyne response of the cavity to periodic laser power switching. Below: time sequences of TTL signals used in the HCRDS experiment to trigger ring-down signals (1), to start the data acquisition process (2) and to determine the frequency switching time of the microwave generator controlling B-EOM (3).

## Section S2. *Ab initio* calculations of line parameters for the HD molecule

Quantum scattering calculations were performed on the potential energy surface (PES) of the $H_2$-$H_2$ system[6] within the Born-Oppenheimer (BO) approximation for the separation of electronic and nuclear motion. The PES is six-dimensional, i.e., it depends on the intermolecular distance, $\tilde{R}$, the three Jacobi angles, $\theta_1$, $\theta_2$, and $\varphi=\varphi_1-\varphi_2$, and the intramolecular distances, $r_1$ and $r_2$ (see Ref.[6] for details). Since the PES is calculated in the BO approximation, it can be used to study all possible combinations of hydrogen isotopologues, provided that the Jacobi angles are transformed accordingly[7-9]. Here, we used the $H_2$-$H_2$ PES to study HD-HD collisions.

To solve the coupled channels equations, the PES was expanded over bispherical harmonics, $I_{l_1 l_2 l_{12}}(\theta_1, \theta_2, \varphi)$

$$V(\tilde{R}, r_1, r_2, \theta_1, \theta_2, \varphi) = \sum_{l_1 l_2 l_{12}} A_{l_1 l_2 l_{12}}(\tilde{R}, r_1, r_2) I_{l_1 l_2 l_{12}}(\theta_1, \theta_2, \varphi), \tag{6}$$

where the bispherical harmonics are defined as

$$I_{l_1 l_2 l_{12}}(\theta_1, \theta_2, \varphi) = \sqrt{\frac{2 l_{12}+1}{4\pi}} \sum_m C^{l_1 l_2 l_{12}}_{m,-m,0} Y_{l_1 m}(\theta_1, \varphi_1) Y_{l_2,-m}(\theta_1, \varphi_2). \tag{7}$$

Here, $C^{l_1 l_2 l_{12}}_{m,-m,0}$ are the Clebsch-Gordan coefficients, and $Y_{lm}(\theta, \varphi)$ denotes the spherical harmonic. Since both collisional partners are heteronuclear, the indices cover both even and odd values, with the restriction that $|l_1 - l_2| \le l_{12} \le l_1 + l_2$, and that the sum $l_1+l_2+l_{12}$ is an even integer. We note that the expansion terms with odd $l_1$ and $l_2$ (which are absent in the expansion of the PES for the $H_2$-$H_2$ system) emerge due to the coordinate transformation (the center-of-mass shift). The sum over $l_1,l_2,l_{12}$ was terminated at the 140th term corresponding to 6,6,12, which ensured the reconstruction of the initial HD-HD PES with a relative root-mean-square error at the level of $1.2 \times 10^{-2}$ %.

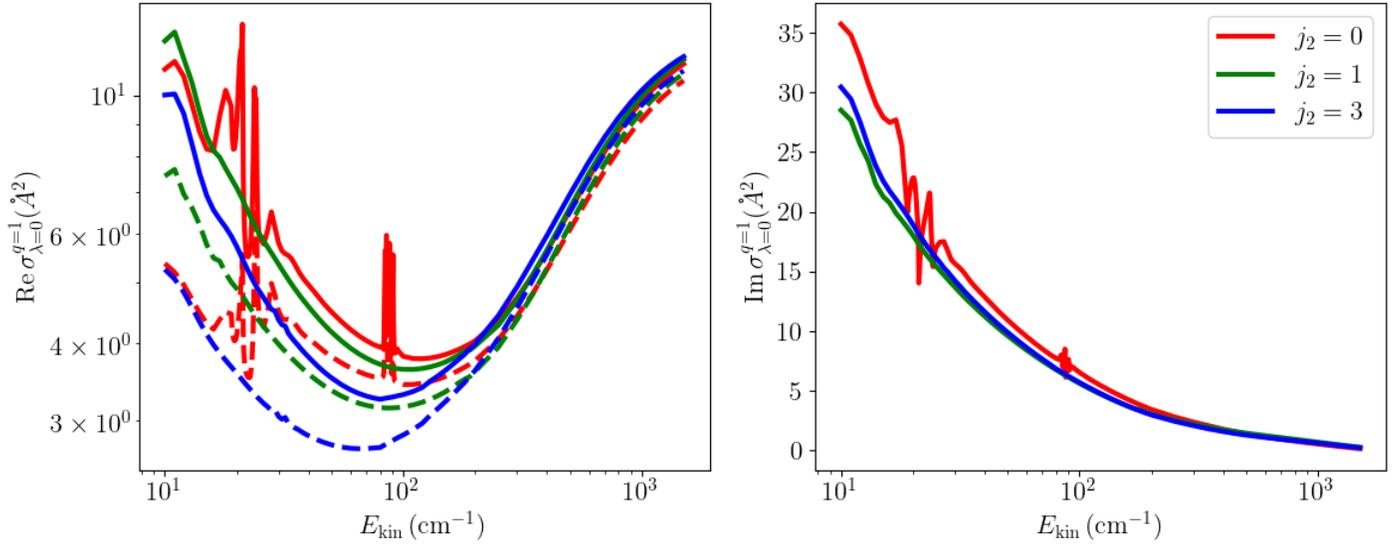

**Figure S3 | Generalized spectroscopic cross-sections.** Pressure broadening (left panel) and pressure shift (right panel) cross sections of the self-perturbed P(3) 2-0 line in HD, for selected rotational quantum numbers of the perturbing molecule. The dashed lines in the left panel correspond to the inelastic contribution to the pressure broadening cross-section.

The expansion coefficients, $A_{l_1 l_2 l_{12}}(\tilde{R}, r_1, r_2)$, were averaged over the internuclear coordinates, $r_1$ and $r_2$, corresponding to the spectroscopically active and perturbing molecules, respectively. The spectroscopically active molecule may be in the ground ($v = 0$) or the second excited ($v = 2$) vibrational states, while the perturbing molecule is always in the ground vibrational state (justified for the experimental temperature of 296 K). Thus, the average was performed by integrating the radial coupling terms with the weight corresponding to the squared modulus of the isolated HD wavefunction in the $v = 0, j = 3$ and $v = 2, j = 2$ rovibrational states (for the average over $r_1$) and $v = 0, j$ (for the average over $r_2$), where $j$ corresponds to the rotational level of the perturbing HD molecule. The wave functions of isolated HD molecules were obtained by solving the rovibrational Schrödinger equation for the HD molecule with the potential energy curve from Ref.[10] using the Discrete Variable Representation - Finite Basis Representation method.

The close-coupling equations were solved in the body-fixed frame[11] using the renormalized Numerov's algorithm for energies in the range $E_{kin} \in \langle 10,1500 \rangle$ cm$^{-1}$ with various steps, to describe the channel-opening effects accurately. At sufficiently large $\tilde{R}$, the log-derivative matrix was transformed to the space-fixed frame, where boundary conditions were imposed, allowing for the recovery of S-matrix elements. The convergence of the scattering S-matrix is ensured by a proper choice of the integration range, propagator step, the size of the rovibrational basis, and the number of partial waves contributing to each scattering event[11]. Calculations were performed using the in-house quantum scattering code BIGOS.

The scattering S-matrix elements were used to calculate the generalized spectroscopic cross-sections, $\sigma_\lambda^q(v_i, j_i, v_f, j_f, j_2; E_{kin})$[11-13]. Here, $q$ is the tensor rank of the spectral transition operator (for the electric dipole transition considered here, $q$=1), that drives the transition from the $v_i = 0, j_i = 3$ to the $v_f = 2, j_f = 2$ state in HD. The symbol $j_2$ denotes the rotational quantum number of the perturbing molecule, and $\lambda$ is the rank of the velocity tensor. For $\lambda = 0$, the real and imaginary parts of $\sigma$ correspond to the standard pressure broadening (PBXS) and shift (PSXS) cross sections, respectively. For $\lambda = 1$, the generalized spectroscopic cross section corresponds to the complex Dicke cross section associated with motion narrowing.

Figure S3 presents the pressure broadening (left panel) and pressure shift (right panel) cross-sections for the self-perturbed 2-0 P(3) line in HD. The PBXS for all considered values of $j_2$ exhibit similar behavior to the one observed in He-perturbed lines of $H_2$[14] and HD[15]. First, values of the cross-sections decrease with increasing kinetic energy, then pass through a minimum in the vicinity of $E_{kin} \approx 100$ cm$^{-1}$, and increase in the high-collision energy regime. The dashed lines in the top left panel of Fig. S3 present the inelastic contribution to collisional broadening, i.e., a half-sum of the total inelastic cross-sections in the initial and final spectroscopic states (see Eq. (10) in Ref.[16]). In contrast to He-perturbed HD lines[15], the inelastic contribution to the broadening of self-perturbed HD lines is much more significant, constituting more than 90% of the broadening at kinetic energies larger than 100 cm$^{-1}$. This is attributed to the presence of inelastic channels absent in atom-molecule collisions, i.e., rotational (de-)excitation of the perturber.

Both the PBXS and PSXS for $j_2 = 0$ exhibit resonant features at kinetic energies in the vicinity of 23 cm$^{-1}$ and 90 cm$^{-1}$. These features are associated with channel-opening effects: at $E_{kin} = 22.8690$ cm$^{-1}$, the $(v_1, j_1, v_2, j_2) \to (v_1', j_1', v_2', j_2') =$

(2,2,0,0) → (2,0,0,2) process becomes energetically accessible. This represents a quasi-resonant transfer of rotational quanta Δj = 2 between HD molecules, with an energy mismatch between the ladders of rotational levels in $v = 0$ and $v = 3$. The second resonant structure is related to the excitation of the perturbing molecule, while the active molecule remains in one of the spectroscopic states (the (0,3,0,0) → (0,3,0,1) and (2,2,0,0) → (2,2,0,1) transitions).

We averaged the $\sigma_0^1$ and $\sigma_1^1$ cross-sections over the Maxwell-Boltzmann distribution of relative kinetic energies and summed the resulting values over the relative populations of HD to calculate the six line-shape parameters used in this work in the quadratic speed-dependent hard-collision profile (qSDHCP) and speed-depenendent billiard-ball profile (SDBBP):

- the speed-averaged collisional broadening and shift

$$\gamma_0 - i\delta_0 = \frac{1}{2\pi c}\frac{1}{k_B T}\langle v_r\rangle \sum_{j_2} p_{j_2}(T) \int_0^\infty dx\, e^{-x} \sigma_0^1(j_2; x), \tag{8}$$

- the speed dependences of collisional broadening and shift

$$\gamma_2 - i\delta_2 = \frac{1}{2\pi c}\frac{1}{k_B T}\frac{\langle v_r\rangle\sqrt{M_a}}{2} e^{-y^2} \sum_{j_2} p_{j_2}(T) \int_0^\infty d\bar{x}\left(2\bar{x}\cosh(2\bar{x}y) - \left(\frac{1}{y}+2y\right)\sinh(2\bar{x}y)\right)\bar{x}^2 e^{-\bar{x}^2} \sigma_0^1(j_2; \bar{x}\bar{v}_p), \tag{9}$$

- the real and imaginary parts of the complex Dicke parameter

$$v_{opt}^r - iv_{opt}^i = \frac{1}{2\pi c}\frac{\langle v_r\rangle M_a}{k_B T}\sum_{j_2} p_{j_2}(T)\int_0^\infty dx\, x e^{-x}\left(\frac{2}{3}x\sigma_1^1 - \sigma_0^1\right) x^2 e^{-x^2}. \tag{10}$$

Here, $\langle v_r\rangle = \sqrt{8k_B T/\pi\mu}$ is the mean relative speed of the colliding pair at a given temperature, $k_B$ is the Boltzmann constant, $\mu$ is the reduced mass of the HD-HD system, and $M_a = m_a/(m_a + m_b)$, $x = E_{kin}/k_B T$, $\bar{x} = v_r/\bar{v}_p$, $y = \sqrt{m_a/p}$, and $m_a$ and $m_p$ are the masses of the active and perturbing molecules, respectively. For the self-perturbed case considered here, the mass of the perturber ($m_p$) and the active molecule ($m_a$) is the same, hence, $M_a = 1/2$. Note that in $\sigma_\lambda^1$ symbols, we kept the quantum numbers related to the spectroscopically active molecule implicit for brevity. The quantity $p_{j_2}(T)$ is the population of the $j_2$ level at a given temperature

$$p_{j_2}(T) = \frac{(2j_2 + 1)e^{-E_{j_2}/k_B T}}{\sum_{j_2'}(2j_2' + 1)e^{-E_{j_2'}/k_B T}}. \tag{11}$$

Summations over $j_2$ were truncated at $j_{2\max} = 3$, which covered 97% of the total population of the HD molecule at the experimental temperature. To account for the remaining 3% of the HD population at $T = 296$ K, the cross-sections were extrapolated for $j_2 > 3$ using values for $j_2 = 3$, justified by the lack of significant $j_2$-dependence of the cross-sections, as seen in Fig. S3. The final values of the line-shape parameters at 296 K are gathered in Table 1. Uncertainties of line-shape parameters were estimated following the procedure described in Sec. 5 of Ref.[14].

**Table 1 | Line-shape parameters calculated for the HD P(3) 2-0 transition.** This table contains *ab initio* line-shape parameters (in 10⁻³ cm⁻¹ atm⁻¹) used in the qSDHCP and SDBBP along with the estimated 1σ standard uncertainties.

| $\gamma_0$ | $\delta_0$ | $v_{opt}^r$ | $v_{opt}^i$ | $\gamma_2$ | $\delta_2$ |
|---|---|---|---|---|---|
| 15.5 ± 1.1 | −7.0 ± 1.1 | 27.5 ± 0.2 | −1.0 ± 1.3 | 4.8 ± 0.4 | 1.16 ± 0.25 |